\pdfoutput=1
 \documentclass[aps, pra, superscriptaddress, twocolumn,
	10pt,
	floatfix, 
    nofootinbib,
	tightenlines
]{revtex4-1}

\usepackage[final]{graphicx}
\usepackage{times,bbm,amsmath,amssymb}
\usepackage{epsfig,color}
\usepackage{xcolor}
\usepackage{hyperref}
\hypersetup{colorlinks = true}
\usepackage{cleveref}
\usepackage{microtype}
\usepackage[normalem]{ulem}
\usepackage{float} 
\usepackage{physics}
\usepackage[caption = false]{subfig}
\usepackage[greek,english]{babel}
\usepackage{thumbpdf,enumerate}
\usepackage{booktabs}
\usepackage{sidecap}
\usepackage[scaled=.8]{couriers}
\usepackage{multirow}
\usepackage{placeins}
\usepackage{relsize}
\usepackage{pst-grad,bm}
\usepackage{epigraph}
\usepackage{comment}
\usepackage{textcomp, gensymb}
\usepackage{longtable}
\usepackage{ulem} 
\usepackage{acronym}
\usepackage[normalem]{ulem}

\DeclareUnicodeCharacter{0301}{\'{e}}

\begin{document}

\address{Dipartimento di Fisica, Sapienza Universit\`{a} di Roma, Piazzale Aldo Moro 5, I-00185 Roma, Italy}

\vspace{10pt}

\title{Rotational-invariant quantum key distribution based on a quantum dot source}

\author{Paolo Barigelli}
\address{Dipartimento di Fisica, Sapienza Universit\`{a} di Roma, Piazzale Aldo Moro 5, I-00185 Roma, Italy}

\author{Francesco Sirovich}
\address{Dipartimento di Fisica, Sapienza Universit\`{a} di Roma, Piazzale Aldo Moro 5, I-00185 Roma, Italy}

\author{Gonzalo Carvacho}
\email[Corresponding author: ]{gonzalo.carvacho@uniroma1.it}
\address{Dipartimento di Fisica, Sapienza Universit\`{a} di Roma, Piazzale Aldo Moro 5, I-00185 Roma, Italy}

\author{Fabio Sciarrino}
\address{Dipartimento di Fisica, Sapienza Universit\`{a} di Roma, Piazzale Aldo Moro 5, I-00185 Roma, Italy}

\begin{abstract}  

Quantum Key Distribution (QKD) is a cutting-edge field that leverages the principles of quantum mechanics to enable secure communication between parties involved. Single-photon quantum emitters offer remarkable on-demand photon emission, near-unitary indistinguishability, and low multiphoton generation, thereby enhancing the performance of QKD protocols. Standard approaches in which the polarization degree-of-freedom is exploited are limited by the precise alignment between the communicating parties. To overcome this obstacle, the Orbital Angular Momentum (OAM) of light represents a suitable candidate for encoding the information, as it allows the implementation of rotational-invariant photonic states that remove the need for a fixed physical reference frame between the communicating parties. Here, we report the implementation of an on-demand, rotational-invariant BB84-QKD protocol achieved by exploiting a bright quantum dot source, active time-to-spatial demultiplexing, and Q-plate devices with a space-variant pattern to encode hybrid photonic states. Our findings suggest a viable direction for the use of rotational-invariant hybrid states in on-demand QKD protocols, potentially enhancing security and robustness in complex operational scenarios.
\end{abstract}
\maketitle 
\section{Introduction}
\label{sec:introduction}
Quantum Key Distribution (QKD) has emerged as a revolutionary approach to secure communication in an increasingly digital world \cite{Pirandola20,Aquina2025}. Using the principles of quantum mechanics, QKD allows two parties, typically named Alice and Bob, to share a secret message with unconditional security \cite{Wolf2021}. The realization of long-distance QKD links has already been demonstrated in fiber links \cite{PhysRevLett.130.210801}, metropolitan networks \cite{Pittaluga2025} and in free-space channels \cite{Li2025}. 
The current implementations exploit sources based on decoy state methods with weak coherent pulses \cite{Liu:10,Castillo:25} and spontaneous parametric down-conversion \cite{Bock:16,Zhan2025}, even if the latter have a probabilistic generation process and a trade-off between the generation rate and the quality of the photons \cite{Kaneda:15}. In recent years, quantum dot photon sources have emerged as cutting-edge technology for QKD \cite{Couteau2023, Alléaume_2004, commdot, progressdot}, since they can produce single photons and entangled pairs on demand with high purity and indistinguishability \cite{Somaschi2016, Senellart2017,Liu2018,Wang2019, Uppu2021,Heindel:23}, improving security, key generation rates, and reducing errors associated with multiphoton emission \cite{PhysRevLett.134.210801}. QKD protocols have been successfully implemented using quantum dot photon sources under laboratory conditions, optical fibers, as well as free-space channels \cite{Takemoto2015, Bozzio2022, FBB1,Intallura_2009, dotentangled, BassoBasset_2023, Morrison2023, Zahidy2024}. Despite significant progress in terrestrial and metropolitan fiber optic networks \cite{Peev_2009, Chen:09, Chen2021, Stucki_2011}, as well as advances in chip-scale photonic devices \cite{Sibson2017, PhysRevX.8.021009}, long-distance quantum communication over fiber requires trusted nodes to extend the range of applications. An alternative involves the use of free-space optical links for terrestrial and satellite-based quantum communication \cite{Bedington2017, Liao2017, PhysRevLett.120.030501, Chen2021b}. These free-space systems can be hybridized with fiber optic infrastructures to significantly extend the overall communication range. \\
In most scenarios, the degree of freedom of polarization is used to store the information, since it is easy to manipulate and remains stable throughout transmission \cite{Zhang:17}. One drawback of polarization-based encoding is the requirement to maintain alignment between the user's reference frames during transmission. Although this poses only a minor challenge for fixed terrestrial links, it can lead to substantial errors in communication with mobile targets, such as satellites \cite{Bedington2017}. Nonetheless various techniques have been developed and implemented to actively compensate for random polarization rotations \cite{https://doi.org/10.1155/2011/254154, Chatterjee2023}; these approaches typically require additional resources thus increasing the overall complexity of the system. A simpler and passive solution is the implementation of rotational-invariant states \cite{dambrosio_complete_2012}.  
In recent years, the Orbital Angular Momentum of light (OAM) has been deeply studied and widely exploited in quantum information protocols \cite{Molina-Terriza2007,entangledVV,PhysRevLett.89.197901, supranoam}. Photons possessing a non-zero OAM are characterized by the azimuthal phase dependence $\exp{il\phi}$, where $l\hbar$ is the amount of OAM carried by each photon, and $l$ is an unbounded integer value representing discrete quantum states. Among the different properties of OAM, the capability of encoding qudit states is crucial for future communication networks \cite{D'Ambrosio2013, PhysRevApplied.11.064058}. In particular, the implementation of hybrid states, i.e., photons carrying OAM and polarization simultaneously, has been demonstrated to allow for rotational-invariant QKD protocols as the collective rotations are intrinsically compensated when exploiting the first-order OAM \cite{dambrosio_complete_2012} as schematically depicted in Fig.1. This property has been tested \cite{PhysRevLett.113.060503} yet no implementation with on-demand single-photons has been performed.\\

In this study, we experimentally demonstrate the feasibility of implementing a rotationally invariant BB84 quantum key distribution protocol, in which single photons are generated on-demand by a quantum dot source. These photons are actively demultiplexed into distinct spatial modes and encoded into hybrid polarization–orbital angular momentum (OAM) states using Q-plate devices. To assess the resilience of the protocol against reference frame misalignment, we implemented a rotational platform at Bob's station, allowing controlled rotation of the measurement station, thus impliying a change on the reference frame between the parties. This setup allows for the emulation of different and independently varied angular orientations between the sender and the receiver. The protocol exhibits strong robustness against such external rotational disturbances, as experimentally validated by the measured Quantum Bit Error Rate (QBER) and fidelity. These results suggest opportunities for advancing secure quantum communication in both complex free-space environments and fiber-based systems capable of supporting OAM propagation.
Furthermore, our experimental results indicate that quantum dot sources exhibit encouraging levels of stability and robustness against external noise and rotational disturbances, highlighting their potential suitability for out-of-laboratory QKD implementations.
\begin{figure}[h!]
\centering
\includegraphics[width=0.5\textwidth]{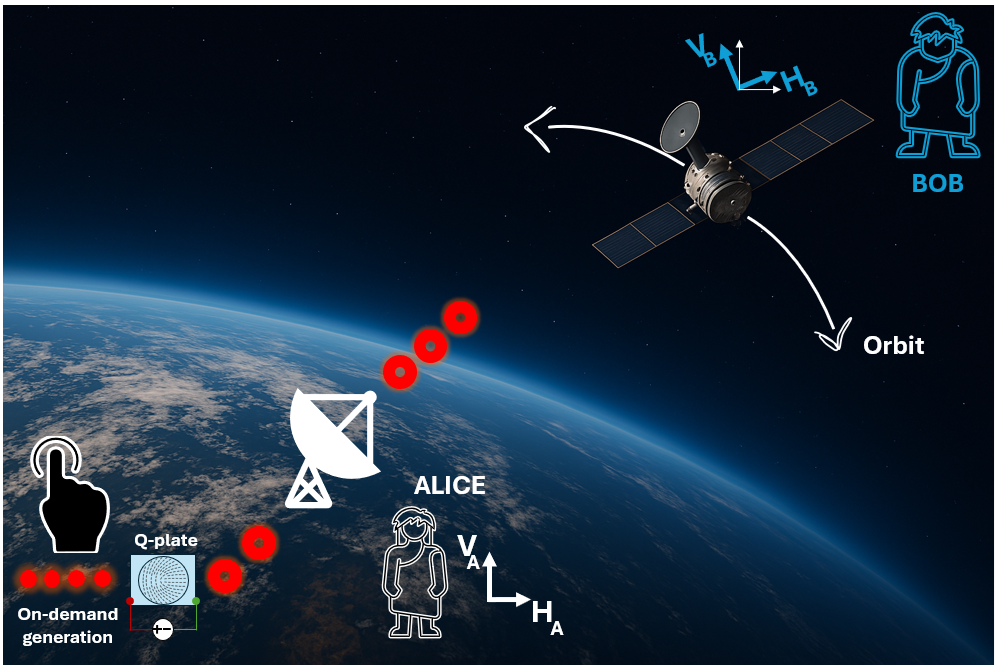}
\label{fig:scheme}
\caption{\textbf{Conceptual scheme of a long-distance, alignment-free quantum communication protocol.} Single photons are generated on-demand and converted into rotational-invariant quantum states using a Q-plate. These states are transmitted to the receiver without requiring reference frame alignment between the communicating parties. An approach that is particularly advantageous when the sender, receiver, or both are in constant motion or subject to rotational fluctuations, which would otherwise hinder the performance of conventional polarization-encoded protocols.}
\end{figure}

\section{Theoretical Framework} 
\label{sec:rot_qkd}
In the BB84 protocol \cite{BB84}, the classical bits are encoded using two mutually unbiased bases: $\mathbb{Z} = \{ \ket{0}, \ket{1} \}$ and $\mathbb{X} = \{ \ket{+}, \ket{-} \}$, where $\mathbb{Z}$ represents an arbitrary orthonormal basis in the qubit Hilbert space, and $\ket{\pm} = (\ket{0} \pm \ket{1})/\sqrt{2}$. Before the transmission, Alice and Bob agree on which quantum states within these bases will represent the classical bits 0 and 1. During each round of the protocol, Alice randomly selects one of the two bases, $\mathbb{Z}$ or $\mathbb{X}$, and prepares a qubit in the appropriate state for sending to Bob. Upon reception, Bob independently and randomly chooses a measurement basis, either $\mathbb{Z}$ or $\mathbb{X}$, to measure the incoming qubit. After many rounds, Alice and Bob publicly announce the basis used in each measurement over a classical channel. They retain only the instances in which their basis choices match, discarding the others. The resulting string of bits from these matching rounds constitutes the so-called sifted key.\\

A rotational-invariant QKD protocol is based on particular states of light that combine the Spin Angular Momentum (SAM) and the Orbital Angular Momentum (OAM) degrees of freedom \cite{dambrosio_complete_2012}. These states are defined as \textit{hybrid states} and are represented as $\ket{\Lambda,m}$, where $\Lambda$ is the polarization component and $\pm m$ is the quantity of OAM carried by the photon in units of $\hbar$ ($m \in \mathbb{Z} $). If we consider the eigenstates $\{\ket{R},\ket{L}\}$ of the SAM operator $\hat{S}_{z}$ and the states $\ket{\pm 1}$ of the OAM operator $\hat{L}_{z}$, where z is the direction of propagation, it is possible to prove that the relative hybrid states are invariant under rotation along the z-axis; in fact, the rotation operator in the SAM (OAM) space is given by the expression $\hat{R}(\theta)=\exp[-\frac{i}{\hbar}\theta\hat{S}_{z}]$ ($\hat{R}(\theta)=\exp[-\frac{i}{\hbar}\theta\hat{L}_{z}]$), acting on the eigenstates by adding a phase: $\hat{R}(\theta)\ket{R/L}=e^{\pm i\theta}\ket{R/L}$ ($\hat{R}(\theta)\ket{\pm m}=e^{\pm im\theta}\ket{\pm m}$). For that reason, the hybrid states $\ket{R,1}$ and $\ket{L,-1}$, and any combination $\alpha\ket{R,1} + \beta \ket{L,-1}$, are invariant under arbitrary rotations along the propagation direction. \\
The preparation of hybrid states can be performed using Q-plate devices \cite{Slussarenko:10,Beijersbergen:94, PhysRevLett.96.163905} which are birefringent plates with a uniform phase delay $\delta$, a topological defect in the center, and a non-uniform transverse optical axis pattern \cite{Marrucci_2011}. The general transformation provided by a Q-plate is defined as follows: 
\begin{equation}
\begin{split}
    QP(\delta)\ket{R, m} &= \cos\frac{\delta}{2}\ket{R,m} + ie^{-2i\alpha_{0}}\sin\frac{\delta}{2}\ket{L,m-2q} \\
    QP(\delta)\ket{L,m} &= \cos\frac{\delta}{2}\ket{L,m} + ie^{2i\alpha_{0}}\sin\frac{\delta}{2}\ket{R,m+2q}.
\end{split}
\end{equation}
The term $\alpha_{0}$ corresponds to the initial angle between the optical axis and the propagation direction. The parameter $\delta$ can be tuned through an external electromagnetic field, reaching maximum conversion at $\delta = \pi$. \\
Starting from any input polarization state $\ket{\psi} = \alpha\ket{R,0} + \beta\ket{L,0} $ with zero angular momentum (i.e. a Gaussian beam), where $|\alpha|^{2}+|\beta|^{2}=1$, a Q-plate with $q =1/2$ can create rotational-invariant states. The transformation of the Q-plate is defined as follows:
\begin{equation}
    QP_{1/2}(\alpha\ket{R,0} + \beta\ket{L,0})= \alpha\ket{L,-1} + \beta\ket{R,1}
\end{equation}
\begin{figure*}[!htpb]
\centering
\includegraphics[width=0.9\textwidth]{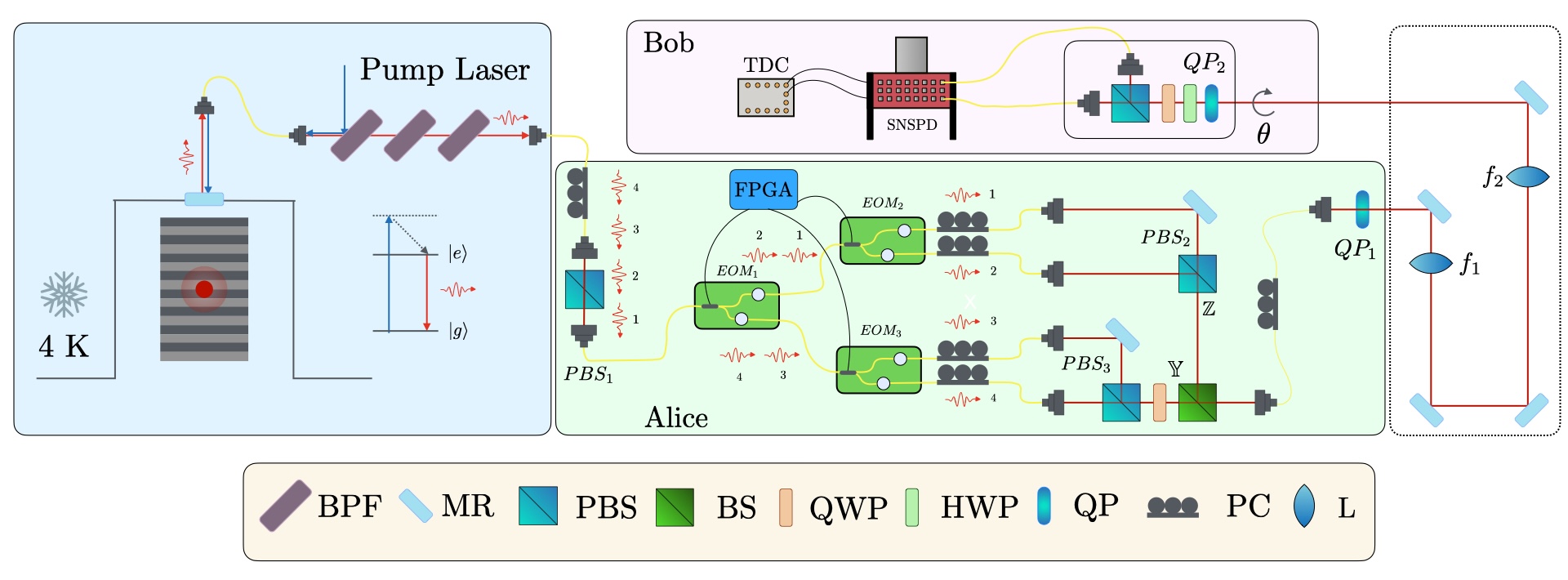}
\label{fig:schematico}
\caption{\textbf{Experimental setup.} Single photons are generated on-demand by a quantum dot, and after the pump laser, they are spectrally cleaned using a series of narrow Band-pass Filters (BPF). The polarization is set to a uniform horizontal state $\ket{H}$ for all photons using a Polarizing Beam Splitter (PBS). The photons are then guided through a Single-Mode Fiber (SMF) to a demultiplexing platform (Green Panel). A sequence of Electro-Optic Modulators (EOMs), controlled by a Field Programmable Gate Array (FPGA), splits the photon stream into four distinct spatial modes by applying synchronized voltage pulses. Each spatial path features a manual Polarization Controller (PC) to fine-tune the polarization before entering the PBS. Photons transmitted through the PBS are prepared in the $\ket{H}$ state, while those reflected are in the $\ket{V}$ state ($\mathbb{Z}$ basis). A Quarter-Wave Plate (QWP) is inserted after one of the PBS outputs to generate circular polarization states in the $\mathbb{Y}$ basis. The four polarization-encoded states are recombined using a $50:50$ Beam Splitter (BS) and coupled into a SMF to ensure spatial mode indistinguishability before interacting with the first Q-plate (QP), where hybrid spin–orbit states are generated. The beam then propagates through a 2-meter free-space channel from Alice to Bob, reflected by mirrors (MR) and resized using a pair of lenses (L) (dotted panel). At Bob’s station (Pink Panel), the received photons encounter a second Q-plate, followed by a Half-Wave Plate (HWP), a QWP, and a PBS—mounted on a rotational platform to enable polarization analysis at arbitrary angles. Finally, the photons are coupled into two SMFs and detected using superconducting nanowire single-photon detectors (SNSPDs). Coincidence counts are recorded via a Time-To-Digital Converter (TDC).} 
\end{figure*}
The application of the same operation via another Q-plate allows for the restoration of the initial input state:
\begin{equation}
QP_{1/2}(\alpha\ket{L,-1} + \beta\ket{R,1}) = \alpha\ket{R,0} + \beta\ket{L,0},    
\end{equation}
Thus, the Q-plate is a tunable device that can perform both the encoding and decoding of the hybrid states.\\
In the experimental implementation, we adopt a hybrid encoding scheme in which the \textit{logical qubits} are defined by:
\begin{equation}
    \begin{aligned}
        \ket{0}_{L} &\equiv \ket{L,-1} \\
        \ket{1}_{L} &\equiv \ket{R,1}
    \end{aligned}
\end{equation}
The relative basis for the QKD protocol is $\mathbb{Y}_{L}$ defined by $\{ \ket{0}_{L}, \ket{1}_{L} \}$, and $\mathbb{Z}_{L}$ equal to $\{\ket{+}_{L}= (\ket{0}_{L} + \ket{1}_{L})/\sqrt{2}, \ket{-}_{L}=(\ket{0}_{L} - \ket{1}_{L})/\sqrt{2} \}$. \\
The hybrid qubit is then transmitted from Alice to Bob. In order to measure it, one first exploits the transformation of a second Q-plate in the measurement stage, and after the transformation of the Q-plate in the measurement stage, the logical qubits are mapped back onto the standard polarization states, resulting in the effective bases $\mathbb{Y} = \{ \ket{R}, \ket{L} \}$ and $\mathbb{Z} = \{ \ket{H}, \ket{V} \}$.
The Quantum Bit Error Rate (QBER) quantifies the fraction of errors in the sifted key. During the QKD protocol, a randomly chosen subset of the sifted key is used to estimate the QBER, and the corresponding bits are subsequently discarded. When using polarization-encoded states, the QBER is sensitive to the relative angular misalignment $\theta$ between Alice's and Bob's reference frames. In fact, the basis $\mathbb{Y}$ does not change with $\theta$, since circular polarizations are invariant under rotation, while $\mathbb{Z}$ adds an error proportional to $\sin^{2}(\theta)$; for that reason, the average QBER varies with $\theta$ according to the relation:
\begin{equation}\label{ref:qber_theta}
QBER(\theta) = \frac{1}{2}\sin^{2}(\theta),
\end{equation}
implying that misalignment can significantly degrade the key quality. In contrast, the hybrid states employed in the present protocol exhibit a QBER that remains constant with respect to $\theta$, as demonstrated experimentally.

\section{Experimental platform} \label{sec:exp_setup}
The experimental platform is depicted in Fig.2. The quantum dot single-photon source (QDSPS) is a commercial Quandela \emph{e-Delight} system. The semiconductor device is made of an InGaAs matrix placed in an electrically controlled micropillar cavity \cite{somaschi2016near} kept at cryogenic temperature (around 4K) by an \emph{Attocube-Attodry800} He-closed cycle cryostat (Fig.2, blue panel). The QDSPS is optically excited by a pulsed laser with a repetition rate of $R_{QD}= 79 MHz$ via the so-called Longitudinal Acoustic (LA) phonon-assisted excitation scheme \cite{PhysRevLett.126.233601}. 
With the excitation laser slightly blue-detuned at approximately $927.8 nm$, filtering of the emitted single photons from the residual laser is achieved using a set of three bandpass filters (BPF) centered at the single-photon wavelength. The source output is characterized by a single-photon rate of $\sim 2 MHz$, measured using Superconductive Nanowire Single-photon Detectors (SNSPDs) \cite{Natarajan_2012}, and a single-photon second-order correlation function of $g^{(2)}(0)\sim 3 \% $, measured through a standard Hanbury-Brown-Twiss (HBT) experiment \cite{brown1956correlation}.\\
The hybrid states are prepared at Alice's station as shown in Fig.2 (Green Panel). 
The photons generated by QDSPS undergo polarization filtering via a Polarizing Beam-Splitter ($PBS_{1}$) before entering an active demultiplexing system. This system, consisting of three Electro-Optic Modulators (EOMs), routes the photons into four separate paths. 
The first electro-optic modulator (EOM$_{1}$) groups the incoming photons into pairs and directs them to EOM$_{2}$ and EOM$_{3}$, which further separate them into four distinct spatial paths. The voltage signals required to activate each EOM at precise time intervals are generated and synchronized by a Field Programmable Gate Array (FPGA). To prepare the polarization states, both the output of $EOM_{2}$ and $EOM_{3}$ feed the input of a bulk Polarizing Beam-Splitter ($PBS_{2}$ and $PBS_{3}$). The output of $PBS_{2}$ consists of the states $\ket{H}$ and $\ket{V}$ and using a Quarter-Wave Plate (QWP) on one output of $PBS_{3}$ we can prepare the states $\ket{R}$ and $\ket{L}$. The four-states ($\ket{H}$, $\ket{V}$, $\ket{R}$ and $\ket{L}$) are recombined on a $50:50$ BS$_{1}$ and coupled into a single-mode fiber to overlap equally the spatial mode of all the polarization states. The fiber transformation applied to the input polarization is compensated using a manual Polarization Controller (PC). In this way, the different beams hit the Q-plate ($QP_{1}$) at the same point, avoiding differences in the conversion efficiency; this enables the generation of rotational-invariant hybrid states without imbalances in their intensity and ensures the same mode distribution after the Q-plate. We implement a telescope to correct the divergence of hybrid states before their transmission; it is made of two lenses with focal lengths of $f_{1}= 50 \ cm  $ and $f_{2} = 40 \ cm $, corresponding to a reducing factor of $M=1.25 $. The reducing factor has been chosen to optimize the Gaussian beam waist at the collimator of the measurement basis.
The hybrid states are propagated in a free-space channel of $\sim 200 cm$ through mirror-assisted reflections.  
The measurement setup at Bob’s terminal (pink panel in Fig.2) consists of a second Q-plate to decode the incoming hybrid states back into polarization qubits, a Quarter-Wave Plate (QWP), a Half-Wave Plate (HWP), and a Polarizing Beam Splitter (PBS), all mounted on a cage rotation system. This configuration allows the entire measurement apparatus to rotate with respect to Alice's reference frame, thereby simulating arbitrary misalignment between their polarization axes. Specifically, measurements were performed at rotation angles of 0 °, 12,5 °, 25 °, 50 °, 75 °, and 90 ° to experimentally validate the resilience of the protocol under varying misalignment of the receiver's reference frame.
At the output branch of the PBS the photons are coupled into single-mode fibers, and are measured by SNSPDs; the detection system exploited has an average detection efficiency of $\eta_{det}\sim 90 \%$ with a dark counts rate of $ R_{dc} < 10 Hz $.\\
\section{Results} \label{results}
We first verify the generation of hybrid quantum states by performing full quantum state tomography \cite{PhysRevA.64.052312} on both the polarization states and the logical ones; for different rotation angles of Bob’s measurement platform $\theta$ ($0^{\circ}$, $12.5^{\circ}$, $25^{\circ}$, $50^{\circ}$, $75^{\circ}$, and $90^{\circ}$), we individually prepared the states $\ket{H}$, $\ket{V}$, $\ket{R}$, and $\ket{L}$ and the states $\ket{0}_{L}$, $\ket{1}_{L}$, $\ket{+}_{L}$, $\ket{-}_{L}$ at Alice's station and transmitted them to Bob. On the receiver side, a complete set of polarization measurements was performed for each input state. The corresponding experimental density matrices were reconstructed using a minimization algorithm based on the maximum likelihood estimation method \cite{PhysRevA.75.042108}.
\begin{figure}[!h]
\centering
\includegraphics[width=0.5\textwidth]{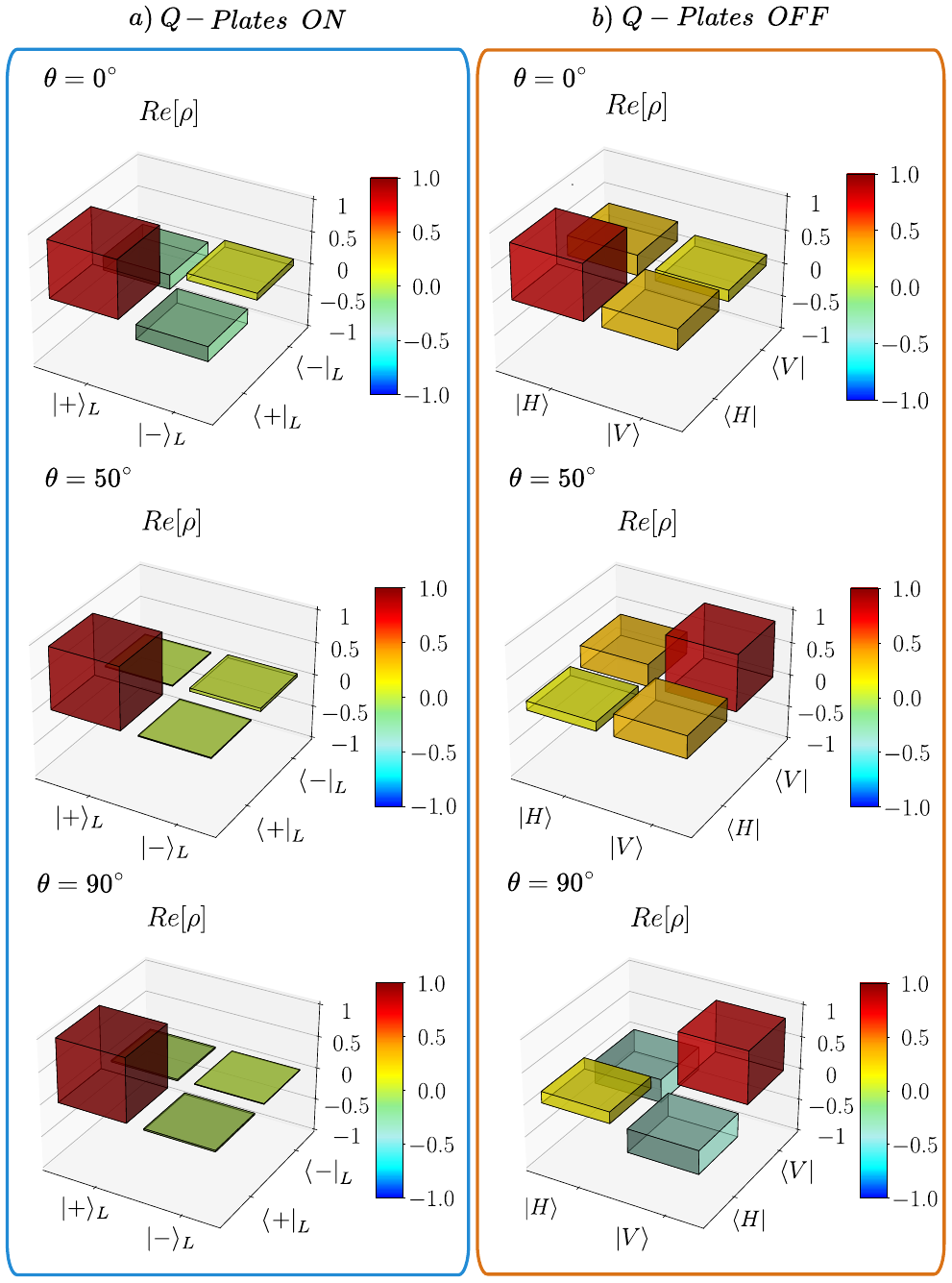}
\label{fig:tomos_H}
\caption{\textbf{Experimental density matrices.} \textbf{a)} Real part of the density matrices for the state $\ket{+}_{L}$ and \textbf{b)} the polarization state $\ket{H}$ at different rotation angles when the Q-plates are active and disabled, respectively. The results demonstrate that, when Q-plates are active, the state remains identical across all angles, confirming its rotational invariance. In contrast, with Q-plates off (polarization-only states), the state evolves as a function of the rotation angle, indicating sensitivity to misalignment.}
\end{figure}
Figure 3-a shows the real part of the experimental states $\ket{+}_{L}$ and Figure 3-b for the state $\ket{H}$ for the angles $0^{\circ}$, $50^{\circ}$ and $90^{\circ}$. The quality of the hybrid state is demonstrated by the invariance of the state $\ket{+}_{L}$ for all angles, while the polarization state at the disabled Q-plates is modified by rotation as predicted by theory (at $\theta=90^{\circ}$ the states go from $\ket{H}$ to $\ket{V}$). Once the robustness of the prepared hybrid states has been demonstrated, we have analyzed the relative fidelities as shown in Fig. 4-a. The blue bars represent the fidelities of the logical states, which remain constant across all angles, while the orange bars correspond to the polarization fidelities, which, as expected, decrease for the $\mathbb{Z}$ basis due to the variation of the reference frame. \\
Given the experimental fidelities $F_{i}$, the expected QBER can be computed as follows: $QBER = \frac{1}{4}\sum_{i}(1-F_{i})$ where the sum is performed on all polarization states. 
To experimentally address the robustness of the setup against rotations, Fig.4-b shows the measured QBER as a function of the rotation angle $\theta$, along with the theoretical prediction (orange curve) for polarization-only encoding, as described by Eq.5. 
\begin{figure}[!h]
\centering
\includegraphics[width=0.5\textwidth]{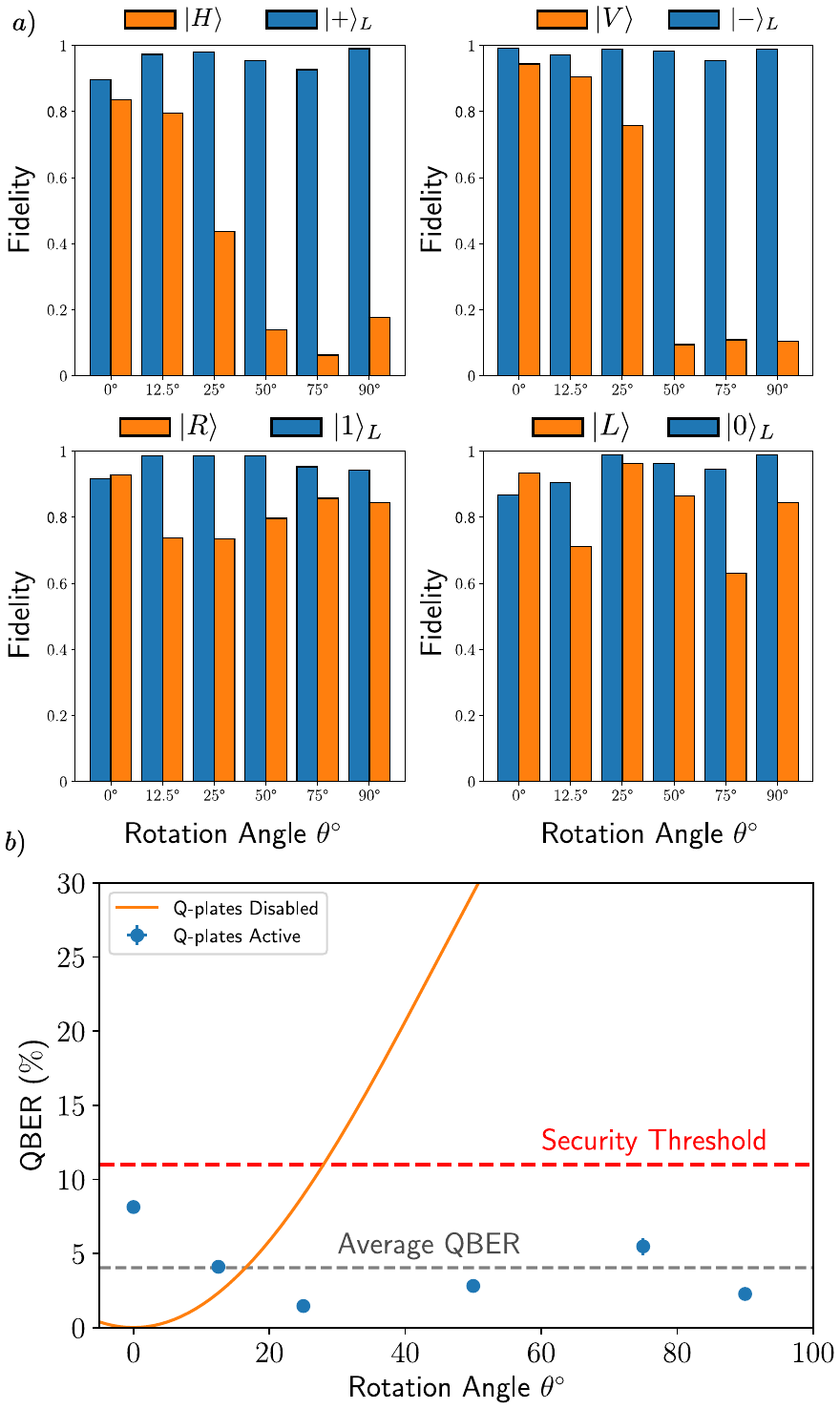}
\label{fig:rotation_QBER}
\caption{\textbf{Experimental fidelities and QBER for different $\theta$ angles between Alice's and Bob's reference frame.} \textbf{a)} Fidelities for the logical (blue bar) and polarization (orange bar) states across different angles $\theta$; in the logical basis, the fidelity is always constant while it decreases for the $\mathbb{Z}$ basis. \textbf{b)} The orange line represents the theoretical QBER in the case that the state is not rotational-invariant ($QBER=\sin^{2}(\theta)/2$). The security threshold depicted as a red dotted line underlines the maximum QBER possible for an effective secret key exchange ($11\%$). The grey line is the average value of the different angles equal to $\langle QBER \rangle = (4.04\pm0.02)\% $. } 
\end{figure}
The experimental QBER remains well below the security threshold of 11\% and is approximately constant in all angles tested, with an average value of $\langle QBER \rangle = (4.04 \pm 0.02)\%$. The observed deviation from the expected sinusoidal behavior of polarization-based systems confirms the rotational invariance of the generated hybrid states. Residual fluctuations in the QBER are attributed to imperfections in the state preparation at Alice’s station rather than sensitivity to rotational misalignment. 
\section{Conclusions}\label{sec:conclusions}
Quantum key distribution (QKD) protocols represent a fundamental building block for securing future communication infrastructures. Enhancing their performance and exploring implementations using cutting-edge technology are therefore essential to address emerging real-world scenarios. In this work, we have experimentally demonstrated the generation of on-demand hybrid quantum states with intrinsic rotational invariance, aimed at overcoming the limitations imposed by alignment requirements between distant communicating parties. The on-demand nature of the photon source is realized through the use of a quantum dot emitting at 925 nm, a wavelength compatible with both free-space channels and integrated photonic platforms. Rotational invariance is achieved by encoding quantum information into hybrid states residing in decoherence-free subspaces, combining polarization and orbital angular momentum (OAM) degrees of freedom. This encoding inherently compensates for arbitrary rotations of the reference frame, eliminating the need for additional alignment mechanisms or feedback systems. We evaluated a key performance metric, the quantum bit error rate (QBER), under two experimental conditions: a stable, aligned configuration representing the ideal case, and a dynamically misaligned scenario simulating practical operational challenges. In both cases, the measured QBER remained consistently below the security threshold, even when continuous rotational misalignments were introduced during key exchange. These results confirm the robustness and integrity of our implementation of the QKD protocol. The integration of a rotational-invariant encoding scheme with a state-of-the-art quantum dot source and active time-to-spatial demultiplexing of single photons enables the mitigation of one of the most critical challenges in the real-world deployment of QKD. The resistance of the hybrid states to external perturbation \cite{Farias2015} makes the protocol a potential candidate for implementation in metropolitan free-space channels. This approach ensures secure and resilient key distribution without requiring strict alignment between sender and receiver. Additionally, the deterministic emission properties of quantum dots with nearly negligible multi-photon events offer a significant advantage over traditional SPDC sources. 

\section*{Acknowledgements}
We thank Danilo Zia for helpful discussions. G.C. acknowledges support from Sapienza Grant n. RM123188F6CDFC61. The authors acknowledge Project ECS 0000024 Rome Technopole, Funded by the European Union and Thales Alenia Space Italia for supporting the Ph.D. fellowship.
\clearpage
\bibliography{biblio}
\clearpage

\end{document}